\documentstyle[epsfig,sprocl]{article}

\def\Journal#1#2#3#4{{#1}~{\bf #2}, #3 (#4)}

\def\NPB{{\em Nucl.~Phys.\/}~B}
\def\NPA{{\em Nucl.~Phys.\/}~A}
\def\PLB{{\em Phys.~Lett.\/}~B}
\def\PRL{\em Phys.~Rev.\ Lett.\/}
\def\PRD{{\em Phys.~Rev.\/}~D}
\def\PRP{{\em Phys.~Rept.\/}}

\def\be{\begin{equation}}
\def\ee{\end{equation}}
\def\bea{\begin{eqnarray}}
\def\eea{\end{eqnarray}}

\begin{document}

\title{COOPER-MESONS IN THE COLOR-FLAVOR-LOCKED SUPERCONDUCTING 
PHASE OF DENSE QCD\,\footnote{FZJ-IKP(TH)-2000-22}}

\author{A. Wirzba} 

\address{Institut f\"ur Kernphysik (Theorie)\\
Forschungszentrum J\"ulich\\
D-52425 J\"ulich, Germany\\
E-mail: a.wirzba@fz-juelich.de}


\maketitle\abstracts{ 
QCD superconductors in the color-flavor-locked (CFL) phase
sustain excitations (``Cooper'' mesons)
that can be described as pairs of particles or holes
around a
gapped Fermi surface. In weak coupling and to leading logarithm
accuracy the masses, decay constants and form factors of the
scalar, pseudoscalar, vector and axial-vector
excitations,  which explicitly are of finite size,
can be calculated exactly. Furthermore, the constraints of this microscopic
calculation  on the effective-lagrangian description and 
the computation of the generalized triangle anomaly are discussed.
}      

\section{Introduction} 
Quantum chromodynamics (QCD) at high density, relevant to the
physics of the early universe, compact stars and relativistic
heavy ion collisions, is presently  attracting a renewed attention
from both nuclear and particle physicists.
For large chemical potential, $\mu \gg \Lambda_{QCD}$, quarks at
the edge of the Fermi  surface interact weakly, although the high
degeneracy of the  Fermi surface causes perturbation theory to fail.
Thus particles can pair as diquarks and condense at the boundary of
the Fermi surface leading to energy gaps and therefore to a
superconducting ground state~\cite{sc70,sc80,sc98,CFL,Son,PiRisch,Reviews}. 
The resulting
QCD superconductor breaks color and flavor symmetry spontaneously.
Therefore, the ground state exhibits Goldstone modes that are either
particle-hole excitations (ordinary pions) or particle-particle and
hole-hole excitations (BCS pions $\tilde \pi$) with a mass that vanishes in the
chiral limit.

Effective-lagrangian approaches to QCD in the color-flavor-locked
(CFL) superconducting phase~\cite{CFL} provide a convenient 
description of the
long-wavelength physics structured by global flavor-color symmetries
(incl.\ anomalies)~\cite{HRZ,GATTO,SONSTE,Zarembo}:
\begin{eqnarray}
{\cal L} &=& 
\frac{F_T^2}{4}{\rm Tr}\left(\partial_0 \tilde U \partial_0 
\tilde U^\dagger\right)
-\frac{F_S^2}{4}{\rm Tr}\left(\partial_i \tilde 
U \partial_i \tilde U^\dagger\right)
+ {\cal L}_{\rm mass}+ {\cal L}_{\rm WZW} + {\cal L}_{\rm h.\, o.}\,,
\label{eq:EffLag}
\end{eqnarray}
where $\tilde U= \exp(i\tau^a \cdot {\tilde \pi}^a/F_T)$ is the pertinent
unitary chiral matrix of the pseudoscalar Goldstone modes $\tilde \pi^a$ 
which are of
diquark nature here.
The nonlinear sigma-model part of this 
effective lagrangian is of the nonrelativistic antiferromagnetic 
type introduced by Leutwyler~\cite{Leutwyler}. Note that the space-like
pion decay constant $F_S$ is smaller than the time-like one $F_T$.
This is a common property for the propagation of 
Goldstone modes in a matter background~\cite{ThorWi,KiWi,PiTyt}.
${\cal L}_{\rm mass}$ is the explicitly symmetry-breaking part,
${\cal L}_{\rm WZW}$ is the (generalized) 
Wess-Zumino-Witten term~\cite{HRZ}, and ${\cal L}_{\rm h.o.}$
stands for all higher order derivative or mass contributions.

However, such an effective-lagrangian approach is 
intrinsically based on a
point-like description and does not allow a direct calculation of the
underlying parameters that are important for a quantitative
description of the bulk (thermodynamic and transport) properties of
the QCD superconductor. This requires a more microscopic description.
In this talk, I report about such a work~\cite{RWZ,RSWZ,NRWZ} where
the excitations  of QCD superconductors in the CFL
phase are described  
microscopically as composite (finite size) pairs of quasi-particles
or quasi-holes  around the gaped Fermi surface. 
The results of microscopic calculations in the leading logarithm approximation
of the masses and decay constants
of these generalized (alias ``Cooper'')
mesons are reviewed. 

\section{QCD Superconductor in the CFL Phase}
In the CFL  phase of the QCD superconductor~\cite{CFL}, 
the quarks are gapped. Their propagation in the
chiral limit is conveniently described 
 in the Nambu-Gorkov formalism where the
fermi field $\psi$ is doubled to $(\psi,\psi_C)$ with
$\psi_C\equiv C \bar\psi^T$.~\cite{PiRisch}
For large chemical potential $\mu$, the antiparticles decouple:
the particle/hole energies are
${{\epsilon}}_q\approx \mp (q_{||}^2+ |G(q)|^2)^{1/2}$,
whereas the anti\-particle/anti\-holes ones are 
$\bar\epsilon_q \approx \mp 2\mu$, with $q_{||}\mbox{=}(|{\bf q}|-\mu)$ 
the momentum measured parallel 
to the Fermi momentum. Thus the quark-propagator 
in the Nambu-Gorkov space reads~\cite{RWZ}
\begin{eqnarray}
  {\bf S}&\approx& \left(\begin{array}{cc}
     \gamma^0\,(q_0+q_{||}) \Lambda^{-}({\bf q})\ \ \
    &  {  -{\bf M}^\dagger\,G^\ast(q)\Lambda^{+}({\bf q})}
  \\
     {{\bf M}\, G(q)\Lambda^{-}({\bf q})\ \  \ } & \gamma^0\,(q_0-q_{||})\
\Lambda^{+}({\bf q})
\end{array}\right)\,\,\frac{1}{q_0^2-\epsilon_q^2} 
 \end{eqnarray}    
in terms of the projectors
 $\Lambda^{\pm}({\bf q})=
{\textstyle\frac{1}{2}}(1\pm {\boldmath\alpha}
\cdot\hat{{\bf q}})$ onto
positive and negative energies and the
color-flavor-locking matrix
${\bf M}_{f_i f_e}^{c_i c_e}= \epsilon_{\ell f_i f_e}\epsilon^{\ell c_i
c_e}\,\gamma_5$ of the CFL phase.
$G(q)$ is the gap {\em function} (not a constant) that 
satisfies the gap equation of
Fig.~\ref{FigGap}:
\begin{figure}[ht]
 \centerline{\epsfig{file=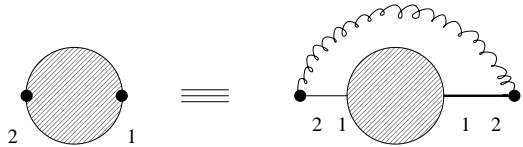,height=1.9cm}}
\caption[]{BCS gap equation. The thin and thick lines are the free
and dressed quark propagator, respectively, whereas the wiggly line
is the gluon propagator. The indices refer to the off-diagonal
components in the Nambu-Gorkov space}\label{FigGap}
\end{figure}   
\begin{eqnarray}
  G(p_{||}) &\approx &
  \frac{h_\ast^2}{6}
\int_0^{\infty}\! dq_{||}\,
   \frac{G(q_{||})}{\sqrt{q_{||}^2+|G(q_{||})|^2}}
\ln\left\{
   \left(1+\frac{\Lambda_{\perp}^2}
             {(p_{||}\mbox{$-$}q_{||})^2+m_E^2}\right)^3 \right.
\nonumber\\[-1mm]
  &&\qquad\mbox{}\times\left.
    \left(1+\frac{\Lambda_{\perp}^3}{|p_{||}\mbox{$-$}q_{||}|^3
 +\frac{\pi}{4}m_D^2|p_{||}\mbox{$-$}q_{||}|}
    \right)^2\,\right\} \; ,
\label{eq:GapEqScreened}
\end{eqnarray}     
where
$
h_*^2=\frac 43 \frac {g^2}{8\pi^2}
$ in terms of the strong coupling constant $g$.\cite{BRWZ}   
The gap equation is not of a conventional BCS-type because of the
long-range structure of the gluon propagator, i.e. 
\begin{equation}
 {\cal D}(q) = \textstyle\frac12 
\displaystyle \frac{1}{q^2 + { m_E^2}}
  + \textstyle\frac12 \displaystyle \frac{1}{q^2 + {m_M^2}}
\label{eq:GluonProp}
\end{equation}
in Euclidean space.
The gluon propagator 
is {\em Debye screened} ( $m_E^2=m_D^2\approx
N_f (g\mu)^2/2\pi^2$) and
{\em Landau damped} ($m_M^2\approx \pi m_D^2{|q_4|/|4{\bf q}|}$),
where
$m_D$ is the Debye mass, $m_M$ is the magnetic screening generated
by Landau damping and $N_f$ the number of flavors.
In weak coupling, the upper limit for the 
transverse momenta is
$\Lambda_{\perp}=2\mu > m_E,m_M$
and the
logarithms in Eq.~(\ref{eq:GapEqScreened}) cannot be expanded.
Instead logarithmic scales $x=\ln(\Lambda_*/q_{||})$ and
$x_0={\rm ln}(\Lambda_*/G_0)$ 
(with 
$\Lambda_*=(4\Lambda_{\perp}^6/\pi m_E^5)$) 
have to be introduced.
To leading logarithm accuracy, the gap function $G(q)$ is then 
solely a real-valued function of $q_{||}$ given by
\cite{Son,BRWZ}
\begin{equation}
 G(x)=G_0\,\sin \left(\frac {\pi x}{2x_0}\right)
 = G_0 \sin \left(h_* x/\sqrt{3}\right)
\quad\mbox{with}\ 
G_0 \approx \left(\frac{4 \Lambda_{\perp}^6}{\pi m_E^5}\right)\,
  e^{-\frac{\sqrt{3}\pi}{2h_\ast}} \, . 
\end{equation}

\section{Generalized Scalar and Pseudoscalar Mesons}
\begin{figure}[hb]
 \centerline{\epsfig{file=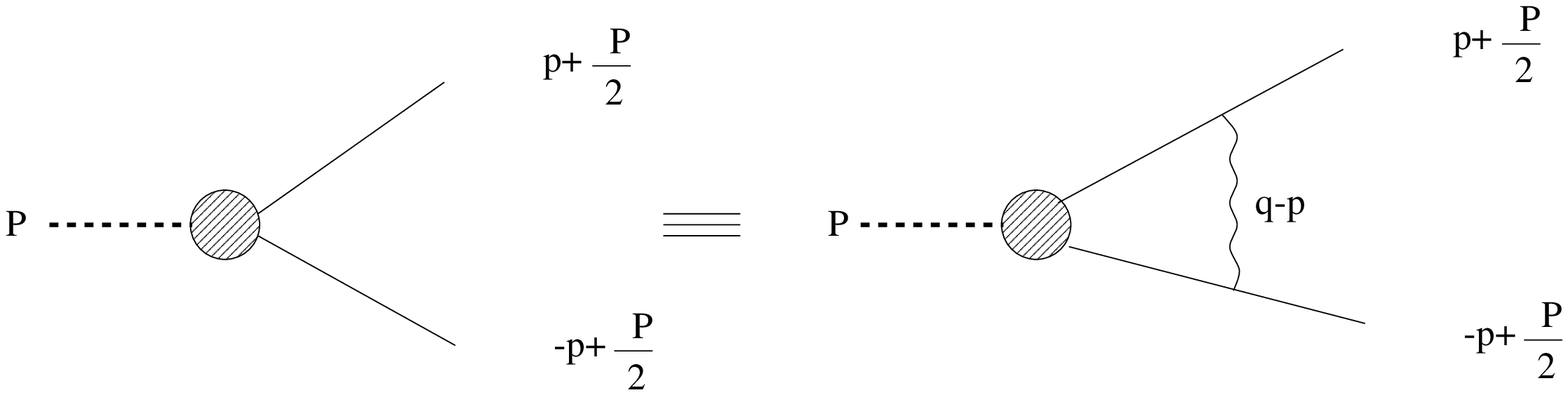,height=2.25cm}}
\caption{Bethe-Salpeter equation for the generalized mesons in the QCD
   superconductor.}
  \label{Fig_BSeq}
\end{figure}

The generalized mesons of the CFL phase 
are excitations in $qq$, whereas the
conventional mesons are excitations in $\bar q q$.
Their wave-functions follow form the Bethe-Salpeter equation pictorially
described
in Fig.~\ref{Fig_BSeq}~\cite{RWZ,RSWZ}.
In particular, in leading logarithm approximation, the scalar
vertex operator of `isospin' $A$ for a pair of particles
or holes with momenta $P/2\pm p$ 
is given in the Nambu-Gorkov space by
\begin{eqnarray}
  {\bf \Gamma}^A_{\sigma}\,(p,P)= \frac {1}{F_T}\,\,
\left(\begin{array}{cc} 0 &  i\Gamma_{S}^\ast(p,P)\,
      \left({\bf M}^{A}\right)^\dagger 
                  \\
            i\Gamma_{S} (p,P)\,{\bf M}^A 
  & 0\end{array}\right)
\label{Svertex}
 \end{eqnarray}
with ${\bf M}^A={\bf M}^{i\alpha}\,( \tau^A)^{i\alpha}$
and ${\bf M}^{i\alpha}=\epsilon_f^i\,\epsilon_c^\alpha\,
\gamma_5$, where
$(\epsilon^a)^{bc}=\epsilon^{abc}$ is the totally antisymmetric tensor
in flavor ($f$) and color ($c$) space, respectively. 
The corresponding composite pseudoscalar vertex reads
\begin{eqnarray}
  {\bf \Gamma}^A_{\pi}\,(p,P)= \frac {1}{F_T}\,\,
\left(\begin{array}{cc} 0 &  -i\gamma_5\,\Gamma_{PS}^\ast(p,P)\,
      \left({\bf M}^{A}\right)^\dagger 
                  \\
            i\gamma_5\,\Gamma_{PS} (p,P)\,{\bf M}^A 
  & 0\end{array}\right)\,.
\label{PSvertex}
 \end{eqnarray}
In the chiral limit, the scalar and pseudoscalar vertex reduces to the gap,
i.e. $\Gamma_{(P)S} (p, 0) = G(p)$. 
Therefore the scalar and pseudoscalar modes are massless Goldstone modes.
In Ref.\,\cite{RWZ} the generalized PCAC relation
\begin{equation}
 \langle {\rm BCS} | {\bf A}_{\mu}^\alpha (0) |
\tilde\pi^\beta_B (P)\rangle \equiv
iF\,P_{\mu}\,\delta^{\alpha \beta}    
\end{equation}
was studied. It was found there (see also Ref.\,\cite{SONSTE}) that
the temporal decay constant of the generalized pion is given by
 $F_T=\mu/\pi$ whereas the spatial one, $F_S$, is smaller by
a factor $1/\sqrt{3}$. Thus the velocity of the Goldstone modes, 
$v=F_S/F_T =1/\sqrt{3}$,  is less than the speed of light ($c\equiv 1$).
The latter property mirrors the propagation of ordinary pions in nuclear
matter~\cite{ThorWi,KiWi,PiTyt}.
The decay constants of the generalized pions are 
 very large compared with the gap
function which is bounded by $G_0$. Thus the Cooper-pions
have a very small (spatial) size (of the order of the inverse momentum
exchanged between quark pairs at the Fermi surface), irrespectively of
any screening. 

The dependence of the mass of the generalized
pion on the current quark mass can be estimated with similar methods
and was studied in Refs.\,\cite{SONSTE,RWZ}.

\section{Higgs Mechanism}

The generalized scalar mesons  
mix with the longitudinal gluons in the CFL phase~\cite{RSWZ}.
In fact, {\em all} the 8 generalized scalars are eaten up by
the longitudinal gluons, such that  the 
8 originally massless gluons (with two
transverse polarizations)  acquire masses (Meissner effect).
This Meissner mass refers to the inverse penetration length of {\em static}
colored magnetic fields in the QCD superconductor which is unexpectedly
small, i.e. $1/g\mu$.~\cite{RSWZ} In weak coupling, the Meissner mass
is of the order of the electric screening mass $m_E\approx g\mu$. It is not
of the order of $g G_0$ as in a conventional superconductor with a constant
(energy independent) gap.
Note, however, that the {\em nonstatic} gluonic modes with energy 
$Q_0> G_0$ sense
``free quarks'' for which there is {\em electric} screening, but no
magnetic screening.

\section{Generalized Vector and Axial-Vector Mesons}
The pairs of (quasi-)quarks or (quasi-)holes at the Fermi surface can
also form generalized vector and axial-vector meson-states in the CFL 
phase~\cite{RSWZ}.
These composites have a similar form factor as the Cooper-pions and
therefore finite size. Because Lorentz invariance is absent, there are
electric and magnetic vector-composites. The difference to 
the (pseudo-)scalar case is the reduced strength of the coupling
in the Bethe-Salpeter (BS) equation. Thus the (pseudo-)scalar BS equation
admits massless modes, while the (axial-)vector one does not.
The mass of the composite vector excitations  is close to (but smaller than) twice
the gap in weak coupling, but goes asymptotically to zero with
increasing coupling thereby realizing Georgi's vector limit in cold
and dense matter~\cite{RSWZ}. Both the
vector and axial-vector Cooper-octets 
are degenerate in leading-log approximation,
in spite of the chiral symmetry breaking of the CFL phase.
Furthermore, it is
shown in Ref.~\cite{RSWZ} 
that the composite vector mesons decouple from the Noether
currents and that they do not decay to pions in leading logarithm
accuracy, contrary to their analogues in the QCD vacuum. Moreover,
they decouple from the gluons and scalars in the CFL phase as well.

\section{Finite Size and Hidden Gauge}
The quark pairs of the mesons  in the QCD superconductor have finite size.
Though the space like separation is of the order $1/\mu$ and therefore
small, the time-like separation is governed by the scale of the
``magnetic mass''
\begin{equation}
 1/m_M\approx 1/(m_E^2G_0)^{1/3}\gg 1/\mu. 
\end{equation}
Hidden gauge symmetry and vector meson dominance (VMD) can therefore be
only approximate concepts and do not hold in weak coupling and to
leading-log approximation~\cite{RSWZ}.

\section{Modified Triangle Anomaly}
In the CFL phase the ordinary photon is screened, and the 
gluons are either screened or higgsed. However, it was pointed out
in Ref.\,~\cite{CFL} that the CFL phase is transparent to a modified
or tilde photon,
\begin{equation}
  \tilde{A}_\mu= A_\mu\, \cos\theta + H_\mu\,\sin\theta\; ,
 \label{eq:TILDE}
\end{equation}
where 
$A_\mu$ is the ordinary photon field,
which couples to the charge matrix $e\,{\bf Q}_{em} 
=e\, {\rm diag}(2/3,-1/3,-1/3)$ of the quarks, and
$H_\mu$ is the gluon field for $U(1)_Y$ where $g\,{\bf Y}
=g\,{\rm diag}(-2/3,1/3,1/3)$ is the
color-hypercharge matrix. Furthermore,
$\cos\theta=g/\sqrt{e^2+g^2}$,  
and $\sin\theta=e/\sqrt{e^2+g^2}$. 
$\tilde{A}_\mu$ carries color-flavor and tags to the
charges of the Goldstone modes. 
The quark coupling to the tilde photon is in units of
$\tilde{e}=e\cos\theta$. As a result, the CFL phase is
characterized by generalized flavor-color anomalies~\cite{HRZ}.
In Ref.\,\cite{NRWZ} it was shown  how
the triangle anomaly emerges from a direct calculation of the
decay of the generalized pion into two generalized photons,
$\tilde\pi^0\rightarrow\tilde\gamma\tilde\gamma$, in the leading
logarithm approximation, 
\begin{figure}
\centerline{\epsfig{file=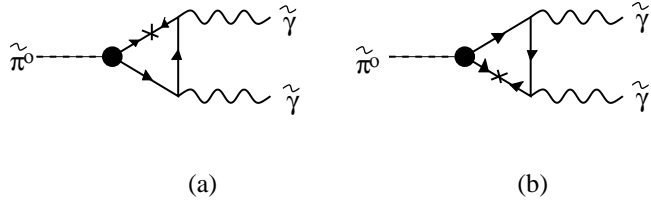,height=1in}}
\caption{Leading contributions to
$\tilde{\pi}^0\rightarrow\tilde\gamma\tilde\gamma$ in the CFL phase.
The crosses refer to pair-condensate insertions.}
\label{Fig_anom}
\end{figure}
i.e.
\begin{eqnarray}
 {\cal T}^{A}_{\mu\nu} (p_1, p_2) =   i \frac{\tilde{e}^2}{F_T}
 \frac{1}{4\pi^2}
 {\rm Tr}\left({\tau^A}\,\widetilde{\bf Q}^2\right)\,
 \epsilon_{\mu\nu\alpha\beta}\,p_1^\alpha\,p_2^\beta 
 \label{eq:PitildeDecay}
\end{eqnarray}
(with $\widetilde {\bf Q} = {\bf Q}_{em}$).
This result is consistent with the modified triangle anomaly 
\begin{equation}
 \partial^\mu\,{\bf A}^3_{\mu} =-\frac{\tilde{e}^2}{96\pi^2}\,
 \epsilon_{\mu\nu\rho\sigma}\,
 \widetilde{F}^{\mu\nu}\,\widetilde{F}^{\rho\sigma}\,\,
\label{eq:ANOMALYCF}
\end{equation}
suggested by the generalized WZW term~\cite{HRZ}.
Here $\widetilde{F}^{\mu\nu}$ is the field strength associated to 
Eq.~(\ref{eq:TILDE}).
Note that the conventional triangle anomaly is larger by
a factor $N_c$. This factor is absent here,
because of the color-flavor locking of the quarks.
Furthermore, the
internal line between the two generalized-photon vertices 
in Fig.~\ref{Fig_anom} 
has to correspond to an antiparticle. Therefore,
the amplitude (\ref{eq:PitildeDecay}) 
does not have an explicit $\mu^2$ dependence which naively
would be expected from the momentum integration around the
Fermi surface.
In fact, because of the dependence on $F_T=\mu/\pi$, 
the radiative decay of the ``Cooper'' pion~(\ref{eq:PitildeDecay}) 
vanishes as $1/\mu$ in dense
matter.

\section{Conclusion}
In this talk, I have reported about the so-called {\em Cooper-mesons}: 
generalized scalar, pseudo\-scalar, vector and
axial-vector particle-particle or hole-hole
excitations in the CFL superconductor of QCD in the weak coupling limit. 
The octet scalar and pseudoscalar excitations
are both {\em massless}, but
only the pseudoscalars survive as Goldstone modes,
while the scalars ones are higgsed
by the gluons leading to the {\em Meissner effect}.        
The vectors and axial
vectors are {\em bound} and {\em degenerate} irrespectively
of their polarization. Their  mass
is less than twice the gap. 
Chiral
symmetry is explicitly 
realized in the vector spectrum in the CFL phase in
leading logarithm approximation, in spite of its breaking in general.
In the CFL superconductor the vector mesons are characterized by 
form
factors that are similar but not identical to those of the generalized
pions. These self-generated form factors provide a natural cutoff to 
regulate the effective calculations at the Fermi surface.           

The composite vector mesons can {\em only} be
viewed as hidden-gauge excitations
if their size is ignored (their form factor set to one).
Only in this approximate limit the
{\em effective lagrangian} description, vector dominance and 
universality are valid. The zero-size
limit is not compatible with the weak-coupling limit, because of
the long-range pairing mechanism at work at large quark chemical
potential. It is an open question whether going beyond the
weak-coupling and leading-log approximations would render the hidden-gauge
or vector-meson-dominance concepts
of effective field theories  more
appropriate. 
The finite-size of the generalized pion, however, 
does not upset the geometrical
normalization of the triangle anomaly 
in the generalized WZW form of Ref.\,\cite{HRZ}.
Much like in the vacuum, the radiative decay of the generalized pions
is dictated by geometry in leading order. Furthermore, it 
vanishes at asymptotic densities.

The existence of {\em bound light}
vector and axial-vector mesons in QCD at high
density  may have interesting
consequences on dilepton and neutrino emissivities
in dense environments: 
e.g., in young and hot neutron stars neutrino production 
via quarks
in the superconducting phase can be substantially modified if the vector
excitations are deeply
bound with a non-vanishing coupling. 
These excitations may be directly seen by scattering electrons off
compressed
nuclear matter
(with densities that allow for a superconducting phase to form)
and may cause substantial soft dilepton emission in the same
energy range in ``cold" heavy-ion collisions.

\section*{Acknowledgments}
It is  
a pleasure for me  to thank my collaborators on this project: Ismail Zahed,
Mannque Rho, Maciej Nowak,  
Byung-Yoon Park, and Edward Shuryak.


\section*{References}
\begin{thebibliography}{99}

\bibitem{sc70}
B.C.~Barrois,
\Journal{\NPB} {129}{390}{1977};
S. Frautschi, Proceedings of workshop on hadronic matter at extreme density,
Erice 1978.
                                                 
\bibitem{sc80}
D.~Bailin and A.~Love,
\Journal{\PRP}{107}{325}{1984}.
  
\bibitem{sc98}
M.~Alford, K.~Rajagopal and F.~Wilczek,
\Journal{\PLB}{422}{247}{1998}; 
R.~Rapp, T.~Sch\"afer, E.V.~Shuryak and M.~Velkovsky,
\Journal{\PRL}{81}{53}{1998}.

\bibitem{CFL} M.~Alford, K.~Rajagopal and F.~Wilczek, 
\Journal{\NPB}{537}{443}{1999}; 
T. Sch\"afer and F. Wilczek, \Journal{\PRL}{82}{3956}{1999}.                  

\bibitem{Son}
D.T.~Son,
\Journal{\PRD}{59}{094019}{1999}.

\bibitem{PiRisch}
R.D.~Pisarski and D.H.~Rischke,
\Journal{\PRD}{60}{094013}{1999}.


\bibitem{Reviews}    
For recent reviews, see K. Rajagopal, to appear
in Proceedings of Quark Matter '99, hep-ph/9908360;
F. Wilczek, to appear in Proceedings of PANIC '99, hep-ph/9908480;
T. Sch\"afer, nucl-th/9911017;
M. Alford, to appear in Proceedings of TMU-Yale, Dec 1999
hep-ph/0003185. 
             
\bibitem{HRZ}
D.K.~Hong, M.~Rho and I.~Zahed,
\Journal{\PLB}{468}{261}{1999}.

\bibitem{GATTO}
R.~Casalbuoni and R.~Gatto,
\Journal{\PLB}{464}{111}{1999}.
 
 
\bibitem{SONSTE}
D.T.~Son and M.A.~Stephanov,
\Journal{\PRD}{61}{074012}{2000}.

\bibitem{Zarembo}
K.~Zarembo,
\Journal{\PRD}{62}{054003}{2000}.

\bibitem{Leutwyler}
H.~Leutwyler,
\Journal{\PRD}{49}{3033}{1994}.

                              
\bibitem{ThorWi}
V.~Thorsson and  A.~Wirzba,
\Journal{\NPA}{589}{633}{1995}
\mbox{[nucl-th/9502003]}; 
A.~Wirzba and V.~Thorsson,  
hep-ph/9502314.


\bibitem{KiWi}
M.~Kirchbach and A.~Wirzba,
\Journal{\NPA}{604}{395}{1996}
\mbox{[nucl-th/9603017]};
hep-ph/9609291;
\Journal{\NPA}{616}{648}{1997}
\mbox{[hep-ph/9701237]}.

\bibitem{PiTyt}
R.D.~Pisarski and M.~Tytgat,
\Journal{\PRD}{54}{2989}{1996}.

\bibitem{RWZ}
M.~Rho,  A.~Wirzba and I.~Zahed,
\Journal{\PLB}{473}{126}{2000} \mbox{[hep-ph/9910550]}.

\bibitem{RSWZ}
M.~Rho, E.~Shuryak, A.~Wirzba and I.~Zahed,
\Journal{\NPA}{676}{273}{2000} 
[hep-ph/0001104].

\bibitem{NRWZ} 
M.~A.~Nowak, M.~Rho, A.~Wirzba and I.~Zahed, 
hep-ph/0007034.

\bibitem{BRWZ} 
B.-Y.~Park, M.~Rho,  A.~Wirzba and I.~Zahed,
\Journal{\PRD}{62}{034015}{2000} [hep-ph/9910347].

\end{thebibliography}
\end{document}